\documentclass[twocolumn,prb,showpacs,aps]{revtex4-1}

\usepackage{amsfonts}
\usepackage{amsmath}
\usepackage{amssymb}
\usepackage{color}
\usepackage{graphicx}
\usepackage{dcolumn}
\usepackage{bm}
\usepackage{indentfirst}
\usepackage{natbib}

\makeatother 
\begin{document}
\title{Yu-Shiba-Rusinov screening of spins in double quantum dots}
\author{K. Grove-Rasmussen$^1$}
\email{k\_grove@fys.ku.dk}
\author{G. Steffensen$^1$}
\author{A. Jellinggaard$^1$}
\author{M. H. Madsen$^1$}
\author{R. $\check{Z}$itko$^{2,3}$}
\author{J. Paaske$^1$}\textbf{}
\author{J. Nyg\aa rd$^1$}

\affiliation{$^1$Center for Quantum Devices \& Nano-Science Center, Niels Bohr Institute,
University of Copenhagen, Universitetsparken 5, 2100~Copenhagen \O,
Denmark}
\affiliation{$^2$Faculty of Mathematics and Physics, University of Ljubljana, Jadranska 19, SI-1000 Ljubljana, Slovenia}
\affiliation{$^3$Jozef Stefan Institute, Jamova 39, SI-1000 Ljubljana, Slovenia}
\maketitle
\date{}

\textbf{A magnetic impurity coupled to a superconductor gives rise to a Yu-Shiba-Rusinov (YSR) state inside the superconducting energy gap~\cite{YuAPS1965,ShibaPTP1968,Rusinov1969}. With increasing exchange coupling the excitation energy of this state eventually crosses zero and the system switches to a YSR groundstate with bound quasiparticles screening the impurity spin by $\hbar/2$~\cite{Satori1992,Bauer2007,ZitkoPRB2011,YaoPRB2014,Hatter2015}. Here we explore InAs nanowire double quantum dots tunnel coupled to a superconductor and demonstrate YSR screening of spin-1/2 and spin-1 states. Gating the double dot through 9 different charge states, we show that the honeycomb pattern of zero-bias conductance peaks, archetypal of double dots coupled to normal leads~\cite{Jeong2001,vanderWiel2002,Chorley2012}, is replaced by lines of zero-energy YSR states. These enclose regions of YSR-screened dot spins displaying distinctive spectral features, and their characteristic shape and topology change markedly with tunnel coupling strengths. We find excellent agreement with a simple zero-bandwidth approximation, and with numerical renormalization group calculations for the two-orbital Anderson model.}

Yu-Shiba-Rusinov states can be imaged in a direct manner by scanning-tunneling spectrocopy of magnetic adatoms on the surface of a superconductor~\cite{Heinricharxiv2017}. Using superconducting tips, high-resolution bias spectroscopy of multiple sub-gap peaks reveals an impressive amount of atomistic details like higher angular momentum scattering channels, crystal-field splitting and magnetic anisotropy~\cite{Yazdani1997,Heinricharxiv2017,Ji2008,Ruby2016,Franke2011,Hatter2015,Ruby2015,Ruby2016,ZitkoPRB2011}. In general, however, it can be an arduous task to model the complex pattern of sub-gap states~\cite{Ji2008,Hatter2015,Ruby2016}, let alone to calculate their precise influence on the conductance~\cite{Ruby2015}.

In contrast, the 'atomic physics' of Coulomb blockaded quantum dots is simple. Changing the back-gate voltage, subsequent levels are filled one-by-one and the different charge states alternate in spin, or Kramers degeneracies for dots with spin-orbit coupling, between singlet and doublet. With normal metal leads, charge states with spin-1/2 exhibit zero-bias Kondo resonances at temperatures below the Kondo temperature, $T\ll T_{K}$, reflecting a Kondo-screened singlet groundstate. If the leads are superconducting with a large BCS gap, $\Delta\gg k_{B}T_{K}$, this resonance is quenched and the groundstate recovers its doublet degeneracy. The system now displays a YSR singlet excitation close to the gap edge, which can be lowered in energy by increasing $k_{B}T_{K}/\Delta$~\cite{PilletNP2010, Deacon2010a, Grove-Rasmussen2009, Lee2014a, ChangPRL2013, Jellinggaard2016}. Close to $k_{B}T_{K}\approx 0.3\Delta$ it crosses zero and becomes the YSR-screened singlet groundstate.

\begin{figure}[t!]
\centering
\includegraphics[width=7.5cm]{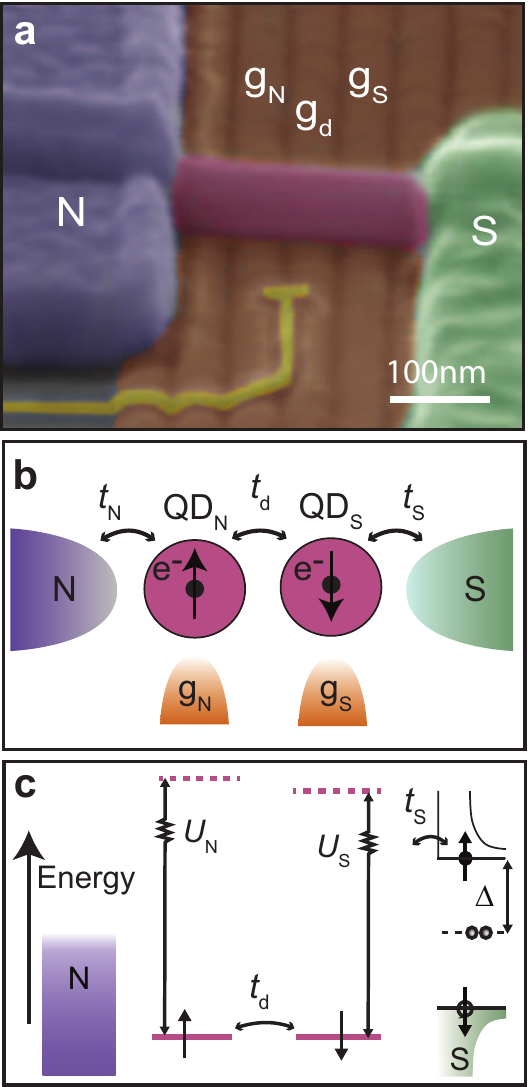}
\caption{Device layout. \textbf{a}, False coloured scanning electron micrograph of Device A showing a normal (N) - InAs nanowire - superconductor (S) device, where a double dot is defined by appropriate voltages on the bottom gates. The yellow floating gate intended for charge sensing by a nearby quantum dot is not used. \textbf{b}, Schematic of a double quantum dot coupled to a normal and a superconducting electrode with couplings $t_\mathrm{N}$ and $t_\mathrm{S}$, respectively. The electrostatic potentials on the two dots are controlled by gates g$_\mathrm{N}$ and g$_\mathrm{S}$, respectively, while the tunnel coupling $t_\mathrm{d}$ between the dots is tuned by gate electrode g$_\mathrm{d}$. \textbf{c}, Energy diagram of a normal - double quantum dot - superconductor device with charging energies larger than the superconducting gap, $U_i\gg \Delta$.}
\label{fig:fig1}
\end{figure}

\begin{figure*}[t!]
\centering
\includegraphics[width=18cm]{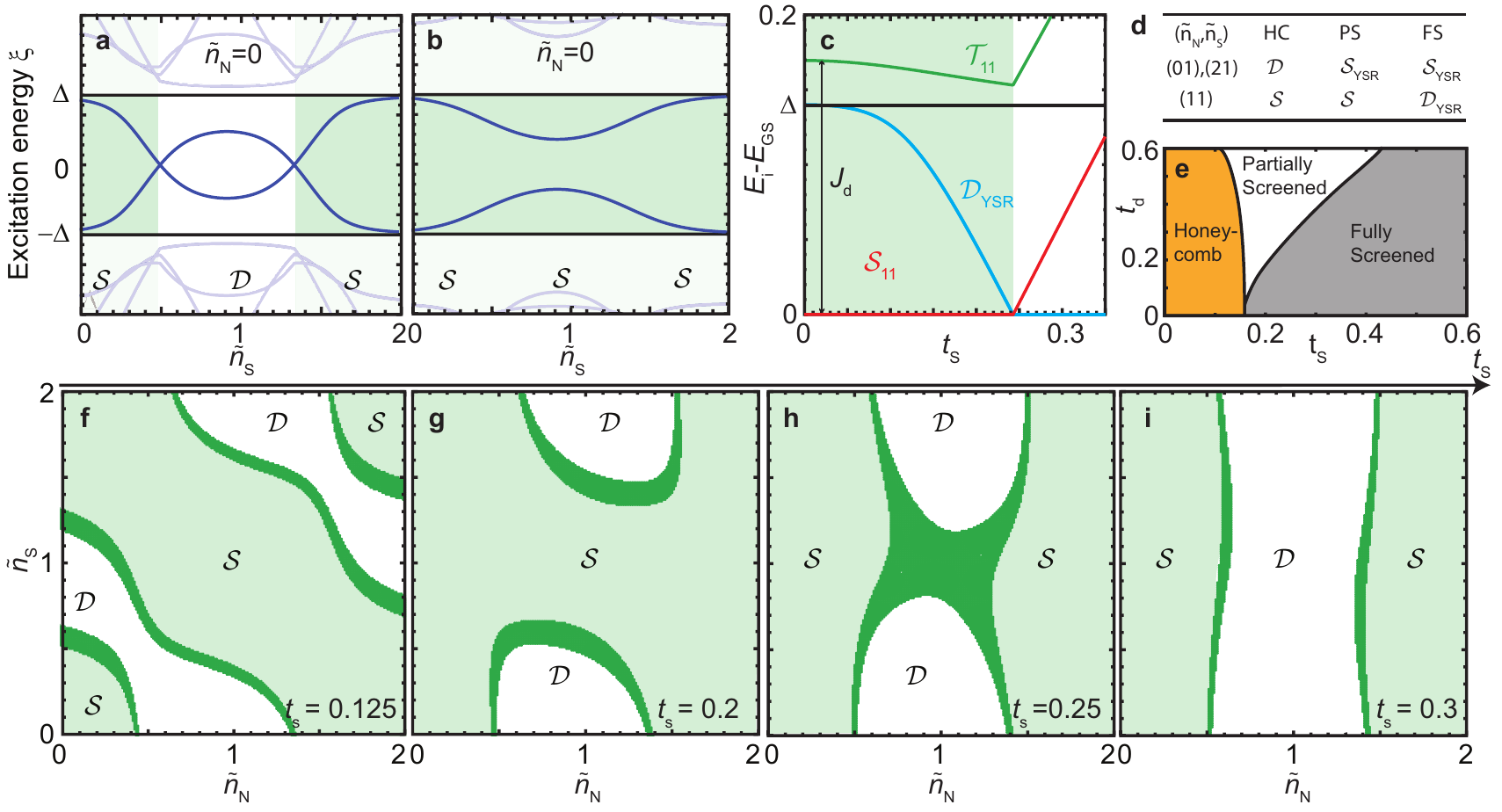}\\
\caption{YSR phase diagrams.
\textbf{a-b}, Two distinct YSR sub-gap spectra vs.\ occupation of QD$_\mathrm{S}$ for $\tilde{n}_\mathrm{N}=0$ (single dot case).
For weak and strong coupling to the superconductor, the doublet (a) and YSR singlet (b) is the ground state for odd occupancy, respectively.
\textbf{c}, ZBW model calculation in the (11) charge state, with singlet $\mathcal{S}_{11}$ and triplet $\mathcal{T}_{11}$ separated by the interdot exchange energy $J_\mathrm{d}$. The triplet is YSR-screened and gives rise to a sub-gap YSR doublet $\mathcal{D}_\mathrm{YSR}$, which becomes the ground state (GS) for large enough $t_\mathrm{S}$. 
\textbf{d-e}, Double-dot YSR phase diagram hosting three phases (e): {\it honeycomb} (HC), {\it partially screened} (PS) and {\it fully screened} (FS). The ground states in the (01) \& (21), and the (11) charge regions for the three regimes are shown in d.
\textbf{f-i}, Stability diagrams for increasing $t_\mathrm{S}$ show the transitions
between honeycomb (f), partially screened (g) and fully screened (i) regimes.
A singlet-doublet energy splitting less than 0.015\,meV (i.e. close to degenerate) defines the dark green region. For \textbf{(h)} $\mathcal{S}_{11}$ and $\mathcal{D}_{\mathrm{YSR}}$ are almost degenerate corresponding to a transition in (11) which is unique to the DQD-S system.
Parameters (in meV) for (d-i): $U_\mathrm{N}=2.5, U_\mathrm{S} = 0.8, U_\mathrm{d} = 0.1, t_\mathrm{d} = 0.27, t_\mathrm{S} = 0.22, \Delta = 0.14$.}
\label{fig:fig2}
\end{figure*}

YSR states were first discussed in the context of gapless superconductivity arising in the presence of randomly distributed paramagnetic
impurities~\cite{YuAPS1965,ShibaPTP1968,Rusinov1969}. However, the ability to assemble spins into dimers, chains and lattices, has recently prompted the exciting idea of engineering YSR molecules~\cite{YaoPRB2014,Kezilebiekearxiv2017}, YSR sub-gap topological superconductors and spiral magnetic states~\cite{Heinricharxiv2017,Nadj-Perge2014,PientkaPRB2013,Schecter2016}. Quantum dots have the advantage of being tunable via electrical gates, and this plays an important role in recent proposals for topological superconductivity in systems of coupled quantum dots~\cite{FulgaNJPh2013,SuNatComm2017,ShermanNatNano2017}. Here we utilize this electrical control to manipulate YSR states in a double quantum dot (DQD) formed in an InAs nanowire. Using multiple finger gates to tune the total DQD spin and the interdot coupling, we demonstrate control of the YSR phase diagram, including electrical tuning between YSR singlets, and a novel YSR doublet arising from the screening of an excited spin triplet.

A scanning electron micrograph of an actual device (Device A) is shown in Fig.~\ref{fig:fig1}a, where bottom gates are used to define a normal-DQD-superconductor structure~\cite{Jellinggaard2016}.
The corresponding schematic is shown in Fig.~\ref{fig:fig1}b, where plunger gates labelled g$_\mathrm{N}$ and g$_\mathrm{S}$ control left (QD$_\mathrm{N}$) and right (QD$_\mathrm{S}$) quantum dot, respectively, while an auxiliary gate, g$_\mathrm{d}$, tunes the interdot tunneling barrier.
The essential physics of this system can be understood in terms of a simple zero-bandwidth (ZBW) model in which the superconductor is modelled by a single quasiparticle coupled directly to an orbital in QD$_\mathrm{S}$ via $t_\mathrm{S}$ and indirectly to QD$_\mathrm{N}$ through $t_\mathrm{d}$. Figure \ref{fig:fig1}c shows the corresponding energy diagram
in the regime of dominating on-site charging energies. In the Supplementary Information (Sec.\ III) we compare this model to NRG calculations to establish its reliability as a quantitative tool.

\begin{figure*}
\centering
\includegraphics[width=18cm]{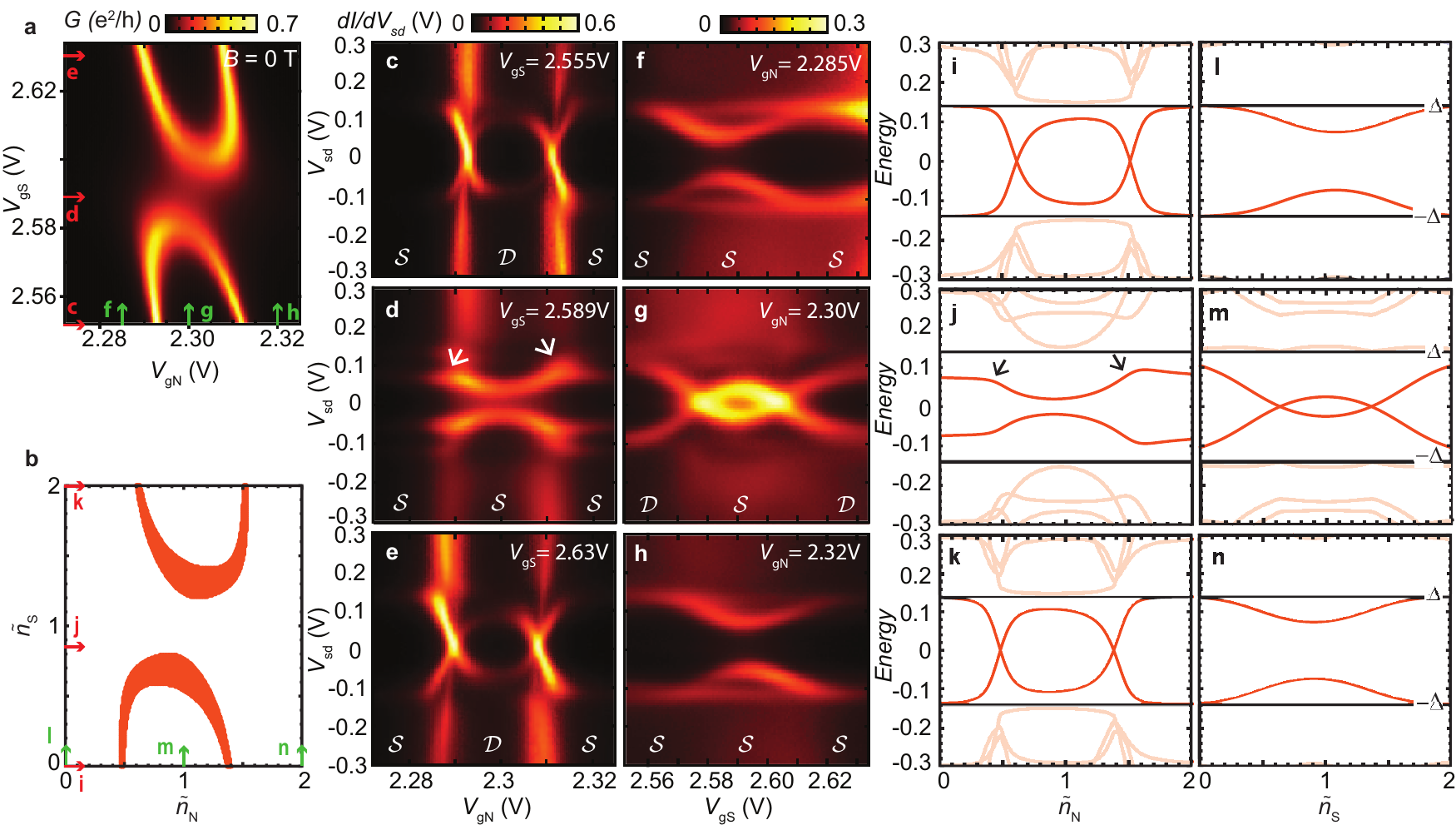}\\
\caption{Sub-gap states in the partially screened regime. \textbf{a}, Stability diagram showing linear N-DQD-S conductance vs. plunger gates at $T= 30$\,mK. \textbf{b}, DQD-S ZBW model reproducing the experimental behaviour in \textbf{a} for intermediate coupling $t_\mathrm{S}$ to S. Orange regions marking ground state transitions have singlet-doublet splitting less than 0.015 meV. \textbf{c-e}, Bias spectroscopy of sub-gap states vs.\ QD$_\mathrm{N}$ occupation, sweeping $V_{\mathrm{g_N}}$ as indicated by red arrows in \textbf{a}. \textbf{f-h}, Sub-gap spectroscopy vs.\ QD$_\mathrm{S}$ occupation sweeping along green arrows in \textbf{a}. All plots show clear sub-gap resonances consistent with the groundstates (doublet $\mathcal{D}$ or singlet $\mathcal{S}$) for different gate voltages indicated below. \textbf{i-n}, ZBW model calculation of sub-gap excitations corresponding to experimental plots \textbf{c-h}. Parameters used in \textbf{b, i-n} are (meV): $U_\mathrm{N} = 2.5, U_\mathrm{S} = 0.8, U_\mathrm{D} = 0.1, t_\mathrm{d} = 0.27, t_\mathrm{S}= 0.22$, and $\Delta = 0.14$. Additional data related to the partially screened regime and how the above parameters are extracted can be found in the Supplementary Information (Sec.\ II.A.)}
\label{fig:fig3}
\end{figure*}

In Fig.~\ref{fig:fig2}a-b, we reproduce the well-known sub-gap state behaviour for a single dot coupled to a superconductor within the ZBW model. 
The panels show excitation energy as a function of the dimensionless gate voltage ${\tilde n}_\mathrm{S}$ (corresponding to the noninteracting average occupation of QD$_\mathrm{S}$) for weak and strong $t_\mathrm{S}$. As expected the sub-gap excitations cross (do not cross) zero energy for weak (strong) coupling. The ground state of the system for odd occupancy is thus a doublet or a YSR singlet (screened spin)~\cite{Satori1992,Deacon2010a,Jellinggaard2016}.

In Figs.~\ref{fig:fig2}f-i we extend the ZBW model to a DQD (finite $t_\mathrm{d}$) and calculate stability diagrams for increasing 
$t_\mathrm{S}$. 
For weak coupling the characteristic honeycomb pattern is observed similar to DQD in the normal state~\cite{vanderWiel2002}. However, as $t_\mathrm{S}$ increases, entirely new types of stability diagrams emerge. In Fig.~\ref{fig:fig2}g, the pattern resembles two mirrored arcs, 
where the lack of zero energy excitations as a function of ${\tilde n}_\mathrm{S}$ for even occupation of QD$_{\mathrm{N}}$ is due to a doublet to singlet transition in QD$_\mathrm{S}$ (see Figs.~\ref{fig:fig2}a,b). Moreover, as the coupling increases even further, the ground state in the $({\tilde n}_\mathrm{N},{\tilde n}_\mathrm{S})=(11)$ region becomes a YSR doublet altering the stability diagram to vertically shaped rectangular regions.

To understand this behaviour, we show the states of the system in the (11) region in Fig.~\ref{fig:fig2}c. Two electrons in the DQD may form either a singlet $\mathcal{S}_{11}$ or a triplet $\mathcal{T}_{11}$ state with energy splitting $J_\mathrm{d}$. Due to the superconductor, a third state may also exist in the gap. In analogy with the QD-S system, where a doublet state may be screened to a YSR singlet, the triplet state may be screened to form a YSR doublet (called $\mathcal{D}_{\mathrm{YSR}}$)~\cite{ZitkoPRB2011}. The energy of $\mathcal{S}_{11}$ and $\mathcal{D}_{\mathrm{YSR}}$ versus coupling is plotted in Fig.~\ref{fig:fig2}c, and the latter eventually becomes the ground state at strong coupling.

The relevant ground states and corresponding phase diagram ($t_\mathrm{S}$ versus $t_\mathrm{d}$) of the DQD-S system with two ground state transitions is shown in Fig.~\ref{fig:fig2}d-e. The first occurs when the system transitions from a honeycomb pattern to the case where the spin in QD$_\mathrm{S}$ is screened. The latter regime we call partially screened (PS) since only some of the charge states are affected. The second transition happens when the $\mathcal{D}_{\mathrm{YSR}}$ in (11) becomes the ground state. This regime we name fully screened (FS) since all possible screened states are ground states of the system.

\begin{figure*}[t!]
\centering
\includegraphics[width=16cm]{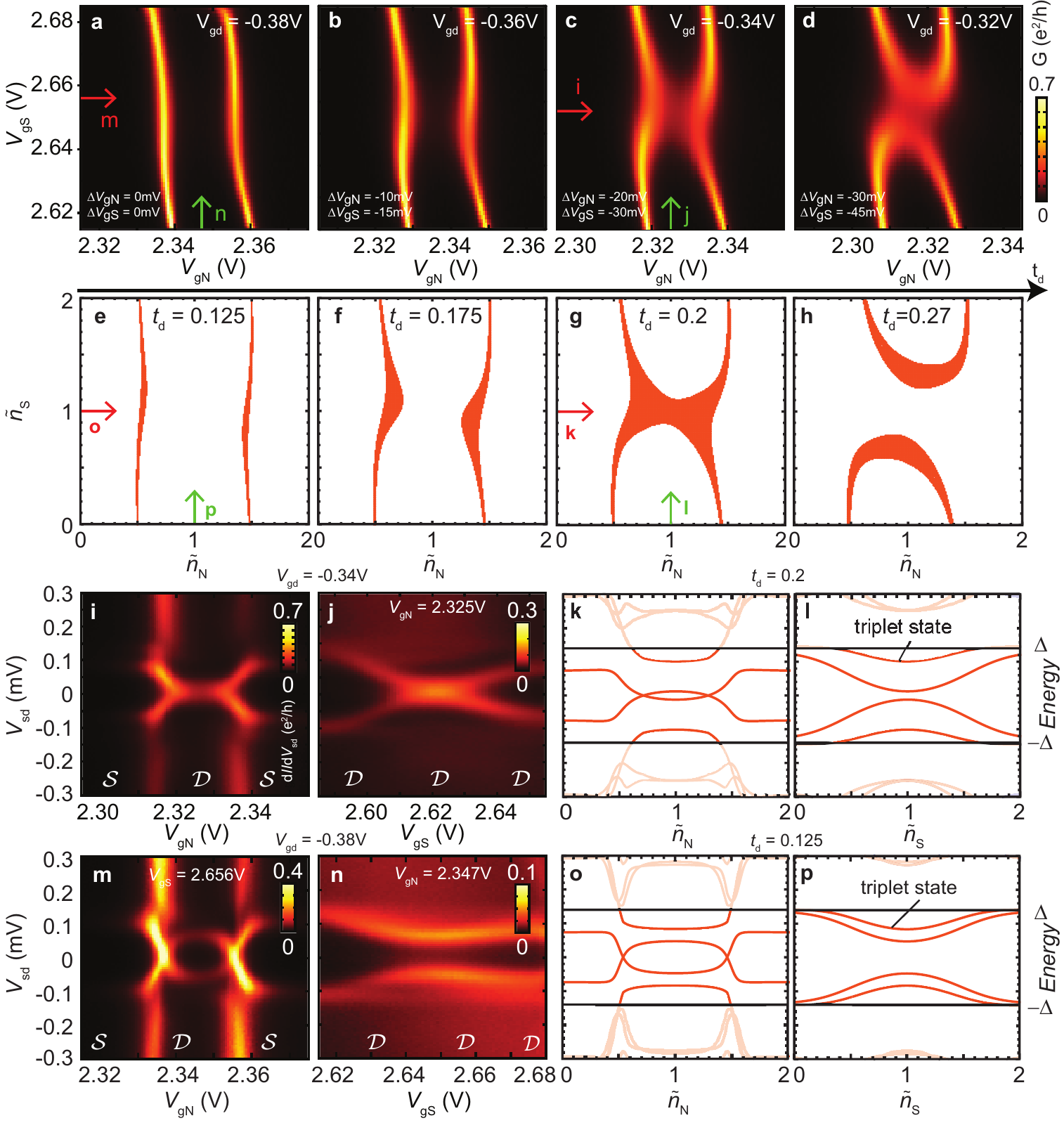}\\
\caption{Tuning interdot coupling $t_\mathrm{d}$. \textbf{a-d}, Stability diagrams for different voltages on the tuning gate g$_\mathrm{d}$ while compensating on plunger gates ($T= 30$\,mK). The plot in \textbf{d} is analyzed in Fig.~\ref{fig:fig3} and the voltage on g$_\mathrm{d}$ is decreased in steps of -20\,mV from \textbf{d} to \textbf{a} (compensating in steps of 10\,mV and -15mV on  g$_\mathrm{N}$ and g$_\mathrm{S}$, respectively). \textbf{e-h},
Stability diagrams generated by the ZBW model for different $t_\mathrm{d}$ (in meV), qualitatively reproducing the experimental behaviour in \textbf{a-d}. Orange regions marking ground state transition have singlet-doublet splitting less than 0.015 meV. \textbf{i-j,m-n}, Bias spectroscopy of sub-gap states vs.\ individual plunger gates swept along red, and green arrows in \textbf{c} and \textbf{a}.  All plots show clear sub-gap resonances consistent with the groundstates (doublet $\mathcal{D}$ or singlet $\mathcal{S}$) for different gate voltages indicated below. \textbf{k-l, o-p}, ZBW model calculation of sub-gap excitations for $t_\mathrm{d} = 0.2$ meV and $t_\mathrm{d} = 0.125$ meV corresponding to experimental plots \textbf{i-j,m-n}. The triplet excitation has very low spectral weight and therefore does not show up in the measured bias spectroscopy. Parameters are fixed to same values as in Fig.~\ref{fig:fig3}. 
}
\label{fig:fig4}
\end{figure*}

With the qualitative behaviour of this system in place, we explore the different regimes experimentally. The honeycomb regime is presented in the Supplementary Information (Sec. II.B, Device B), while below we focus on the stronger coupled regimes. Figure~\ref{fig:fig3}a shows linear conductance versus plunger gates for a two-orbital DQD shell (i.e.\ one spin-degenerate level in each dot). 
A pattern of two arcs are observed resembling the partially screened regime. To verify that the conductance resonances originate from sub-gap states, gate traces for different fillings of the two dots are measured. Figures~\ref{fig:fig3}c-e trace out the filling of electrons in QD$_\mathrm{N}$ along the red arrows in Fig.~\ref{fig:fig3}a, keeping the electron number in QD$_\mathrm{S}$ constant. The sub-gap spectroscopy plots c,e for even filling of QD$_\mathrm{S}$ show similar behaviour, differing from d with odd occupation. When fixing (sweeping) the occupation of QD$_\mathrm{N}$ (QD$_\mathrm{S}$), the qualitative behaviour is switched (Figs.~\ref{fig:fig3}f-h corresponding to green arrows in Fig.~\ref{fig:fig3}a). For even occupancy in QD$_\mathrm{N}$ (f,h) no zero-bias crossing is observed, while the opposite is true for odd occupancy (g). In particular Figs.~\ref{fig:fig3}d,g are interesting, since they involve the (11) charge state region. In contrast to single dot systems, the singlet ground state shows different behaviour whether tuning the electrochemical potential of the dot close to the superconductor or the normal lead, i.e. concave and convex excitation behaviour versus gate voltage in the (11) state. The experimental data clearly confirm that the resonances in the stability diagram originate from sub-gap excitations. The stability diagram generated by our DQD-S model for realistic parameters reproducing the experimental behaviour is shown in Fig.~\ref{fig:fig3}b, and corresponding gate traces for fixed occupations are shown in Figs.~\ref{fig:fig3}i-n. The qualitative agreement between theory and experiment is striking and even subtleties like the asymmetry of the sub-gap resonance splitting in j (see arrows) are reproduced.

The transition between different YSR states can also be driven by changing the singlet-triplet ($\mathcal{S}_{11}$-$\mathcal{T}_{11}$) splitting by tuning $t_\mathrm{d}$  (cf. Fig.\ \ref{fig:fig2}e). In Fig.~\ref{fig:fig4}e-h we show calculated diagrams for $t_\mathrm{d}$ in a parameter range, where the ground state in the (11) charge state transitions from $\mathcal{S}_{11}$ (h) to $\mathcal{D}  _{\mathrm{YSR}}$ (e), i.e. from double-arc, to vertical-lines diagrams. The corresponding measured stability diagrams for the two orbitals analyzed in Fig.~\ref{fig:fig3} are shown in Fig. ~\ref{fig:fig4}a-d, where the gate voltage between the two dots are tuned to more negative values (decreasing $t_\mathrm{d}$). The effect of this tuning qualitatively follows the expectation of the model: a transition from $\mathcal{S}_{11}$ to $\mathcal{D}_{\mathrm{YSR}}$ in the fully screened regime where all spin states are YSR screened.

The gate-dispersion of sub-gap excitations also shows good overall correspondence between ZBW modeling and experiment. We measured the sub-gap spectra in the fully screened regime along the red and green arrows in c and a. The first case c is almost at the transition, where the singlet and doublet states are degenerate in (11). Figures~\ref{fig:fig4}i,j show sub-gap states versus $V_{\mathrm{gN}}$ and $V_{\mathrm{gS}}$, respectively, with a zero bias peak at $V_{\mathrm{gS}}=2.62$\,V reflecting a degeneracy at this value of t$_\mathrm{d}$. The corresponding ZBW modelling in Figs.~\ref{fig:fig4}k,l (for $t_\mathrm{d}=0.25$\,meV) places the system just barely in the fully screened regime with a $\mathcal{D}_{\mathrm{YSR}}$ (11) groundstate and a nearby $\mathcal{S}_{11}$ excitation dispersing very much like in the measurement. A $\mathcal{T}_{11}$ triplet state is predicted inside the gap, and should be accessible from the $\mathcal{D}_{\mathrm{YSR}}$ groundstate. As demonstrated in the Supplementary Information (Sec.\ III.A.), this is confirmed by more accurate NRG calculations, which however reveal a strong suppression of spectral weight on this state, explaining why it may be difficult to observe in experiment. For even lower $t_\mathrm{d}$, case a, Figures~\ref{fig:fig4}m-p again show good overall correspondence between experiment and theory, except for the triplet state, which should be weak, and in this case hardly resolved within the linewidth broadening in the data. Future experiments with hard gap superconductors or improved resolution may eventually lead to capability to detect even such low-weight spectral features. A detailed discussion of nonlinear conductance and broadening of the YSR sub-gap spectra is provided in the Supplementary Information (Sec.\ IV).
\\

\section*{Methods}
The devices are made by defining bottom gate Au/Ti (12/5\,nm) electrodes (pitch 55\,nm) on a silicon substrate capped with 500 nm SiO$_2$ followed by atomic layer deposition of 3x8\,nm HfO$_2$. InAs nanowires (70 nm in diameter) appropriately aligned on bottom gate structures are contacted by Au/Ti (90/5\,nm) normal and Al/Ti (95/5\,nm) superconducting electrodes separated by $\sim 350$\,nm. The superconducting film has a critical field of around 85\,mT. More details on device fabrication can be found in Ref.~\onlinecite{Jellinggaard2016}. The samples are mounted in an Oxford Instruments Triton 200 dilution refrigerator with base temperature of around 30\,mK and are measured with standard lockin techniques. For the data (Fig.\ \ref{fig:fig3},\ref{fig:fig4}) shown in the partially screened regime, the voltages on the gates define a double dot potential with values (V) $V_\mathrm{g1}= 0$, $V_\mathrm{g2}= -1.3$, $V_\mathrm{g3}= 2.3$, $V_\mathrm{g4}= -0.3$, $V_\mathrm{g5}= 2.65$, $V_\mathrm{g6}= -0.3$, and $V_\mathrm{g7}= 0.3$. Here the gate numbers correspond to gates from left to right in Fig.\ \ref{fig:fig1}a. Gates 3, 4, and 5 are thus the left plunger, the tunnel barrier and right plunger gates, which are tuned within some range of the values stated.

\section*{Acknowledgements}
We would like to thank M.C.Hels. Research was supported by the Center for Quantum Devices, The Danish National Research
Foundation, Carlsberg Foundation, The Independent Research Fund Denmark (Natural Sciences), the FP7 FET-Open SE2ND  project and the Slovenian Research Agency (ARRS) under Program P1-0044 and J1-7259.

\section*{Author contributions}
K.G-R.\ and A.J. made the measurements, A.J.\ fabricated the devices and M.H.M.\ grew the nanowires. K.G-R., A.J. and J.N. designed the double quantum dot experiments, G.S., K.G-R.\ and J.P.\ made the ZBW analysis, G.S.\ and J.P.\ the conductance asymmetry calculations, and R.Z.\ provided the NRG analysis. K.G-R., G.S., J.P., R.Z.\ and J.N.\ participated in discussions, analysis and wrote the paper.

\bibliographystyle{naturemag}
\bibliography{arxiv}

\begin{thebibliography}{10}
\expandafter\ifx\csname url\endcsname\relax
  \def\url#1{\texttt{#1}}\fi
\expandafter\ifx\csname urlprefix\endcsname\relax\def\urlprefix{URL }\fi
\providecommand{\bibinfo}[2]{#2}
\providecommand{\eprint}[2][]{\url{#2}}

\bibitem{YuAPS1965}
\bibinfo{author}{Yu, L.}
\newblock \bibinfo{title}{Bound {State} in {Superconductors} with
  {Paramagnetic} {Impurities}}.
\newblock \emph{\bibinfo{journal}{Acta Phys. Sin.}}
  \textbf{\bibinfo{volume}{21}}, \bibinfo{pages}{75} (\bibinfo{year}{1965}).

\bibitem{ShibaPTP1968}
\bibinfo{author}{Shiba, H.}
\newblock \bibinfo{title}{{Classical Spins in Superconductors}}.
\newblock \emph{\bibinfo{journal}{Prog. Theor. Phys.}}
  \textbf{\bibinfo{volume}{40}}, \bibinfo{pages}{435--451}
  (\bibinfo{year}{1968}).

\bibitem{Rusinov1969}
\bibinfo{author}{Rusinov, A.~I.}
\newblock \bibinfo{title}{Superconductivity near a paramagnetic impurity}.
\newblock \emph{\bibinfo{journal}{JETP Lett.}} \textbf{\bibinfo{volume}{9}},
  \bibinfo{pages}{85--87} (\bibinfo{year}{1969}).
\newblock \bibinfo{note}{[Zh. Eksp. Teor. Fiz. {\bf 9}, 146 (1968)]}.

\bibitem{Satori1992}
\bibinfo{author}{Satori, K.}, \bibinfo{author}{Shiba, H.},
  \bibinfo{author}{Sakai, O.} \& \bibinfo{author}{Shimizu, Y.}
\newblock \bibinfo{title}{{Numerical Renormalization Group Study of Magnetic
  Impurities in Superconductors}}.
\newblock \emph{\bibinfo{journal}{J. Phys. Soc. Japan}}
  \textbf{\bibinfo{volume}{61}}, \bibinfo{pages}{3239--3254}
  (\bibinfo{year}{1992}).

\bibitem{Bauer2007}
\bibinfo{author}{Bauer, J.}, \bibinfo{author}{Oguri, A.} \&
  \bibinfo{author}{Hewson, A.~C.}
\newblock \bibinfo{title}{Spectral properties of locally correlated electrons
  in a {Bardeen}-{Cooper}-{Schrieffer} superconductor}.
\newblock \emph{\bibinfo{journal}{J. Phys.: Cond. Mat.}}
  \textbf{\bibinfo{volume}{19}}, \bibinfo{pages}{486211}
  (\bibinfo{year}{2007}).

\bibitem{ZitkoPRB2011}
\bibinfo{author}{\ifmmode~\check{z}\else \v{Z}\fi{}itko, R.},
  \bibinfo{author}{Bodensiek, O.} \& \bibinfo{author}{Pruschke, T.}
\newblock \bibinfo{title}{Effects of magnetic anisotropy on the subgap
  excitations induced by quantum impurities in a superconducting host}.
\newblock \emph{\bibinfo{journal}{Phys. Rev. B}} \textbf{\bibinfo{volume}{83}},
  \bibinfo{pages}{054512} (\bibinfo{year}{2011}).

\bibitem{YaoPRB2014}
\bibinfo{author}{Yao, N.~Y.} \emph{et~al.}
\newblock \bibinfo{title}{Phase diagram and excitations of a {Shiba} molecule}.
\newblock \emph{\bibinfo{journal}{Phys. Rev. B}} \textbf{\bibinfo{volume}{90}},
  \bibinfo{pages}{241108} (\bibinfo{year}{2014}).

\bibitem{Hatter2015}
\bibinfo{author}{Hatter, N.}, \bibinfo{author}{Heinrich, B.~W.},
  \bibinfo{author}{Ruby, M.}, \bibinfo{author}{Pascual, J.~I.} \&
  \bibinfo{author}{Franke, K.~J.}
\newblock \bibinfo{title}{Magnetic anisotropy in {Shiba} bound states across a
  quantum phase transition}.
\newblock \emph{\bibinfo{journal}{Nat. Commun.}} \textbf{\bibinfo{volume}{6}},
  \bibinfo{pages}{8988} (\bibinfo{year}{2015}).

\bibitem{Jeong2001}
\bibinfo{author}{Jeong, H.}, \bibinfo{author}{Chang, A.~M.} \&
  \bibinfo{author}{Melloch, M.~R.}
\newblock \bibinfo{title}{The {Kondo} {Effect} in an {Artificial} {Quantum}
  {Dot} {Molecule}}.
\newblock \emph{\bibinfo{journal}{Science}} \textbf{\bibinfo{volume}{293}},
  \bibinfo{pages}{2221--2223} (\bibinfo{year}{2001}).

\bibitem{vanderWiel2002}
\bibinfo{author}{{van der Wiel}, W.~G.} \emph{et~al.}
\newblock \bibinfo{title}{{Electron transport through double quantum dots}}.
\newblock \emph{\bibinfo{journal}{Rev. Mod. Phys.}}
  \textbf{\bibinfo{volume}{75}}, \bibinfo{pages}{1--22} (\bibinfo{year}{2002}).

\bibitem{Chorley2012}
\bibinfo{author}{Chorley, S.~J.} \emph{et~al.}
\newblock \bibinfo{title}{Tunable {Kondo} physics in a carbon nanotube double
  quantum dot}.
\newblock \emph{\bibinfo{journal}{Phys. Rev. Lett.}}
  \textbf{\bibinfo{volume}{109}}, \bibinfo{pages}{156804}
  (\bibinfo{year}{2012}).

\bibitem{Heinricharxiv2017}
\bibinfo{author}{Heinrich, B.~W.}, \bibinfo{author}{Pascual, J.~I.} \&
  \bibinfo{author}{Franke, K.~J.}
\newblock \bibinfo{title}{{Single magnetic adsorbates on s-wave
  superconductors}}  (\bibinfo{year}{2017}).
\newblock \eprint{ArXiv1705.03672}.

\bibitem{Yazdani1997}
\bibinfo{author}{Yazdani, A.}, \bibinfo{author}{Jones, B.~A.},
  \bibinfo{author}{Lutz, C.~P.}, \bibinfo{author}{Crommie, M.~F.} \&
  \bibinfo{author}{Eigler, D.~M.}
\newblock \bibinfo{title}{Probing the local effects of magnetic impurities on
  superconductivity}.
\newblock \emph{\bibinfo{journal}{Science}} \textbf{\bibinfo{volume}{275}},
  \bibinfo{pages}{1767--1770} (\bibinfo{year}{1997}).

\bibitem{Ji2008}
\bibinfo{author}{Ji, S.-H.} \emph{et~al.}
\newblock \bibinfo{title}{High-resolution scanning tunneling spectroscopy of
  magnetic impurity induced bound states in the superconducting gap of {Pb}
  thin films}.
\newblock \emph{\bibinfo{journal}{Phys. Rev. Lett.}}
  \textbf{\bibinfo{volume}{100}}, \bibinfo{pages}{226801}
  (\bibinfo{year}{2008}).

\bibitem{Ruby2016}
\bibinfo{author}{Ruby, M.}, \bibinfo{author}{Peng, Y.}, \bibinfo{author}{von
  Oppen, F.}, \bibinfo{author}{Heinrich, B.~W.} \& \bibinfo{author}{Franke,
  K.~J.}
\newblock \bibinfo{title}{Orbital {Picture} of {Yu}-{Shiba}-{Rusinov}
  {Multiplets}}.
\newblock \emph{\bibinfo{journal}{Phys. Rev. Lett.}}
  \textbf{\bibinfo{volume}{117}}, \bibinfo{pages}{186801}
  (\bibinfo{year}{2016}).

\bibitem{Franke2011}
\bibinfo{author}{{Franke}, K.~J.}, \bibinfo{author}{{Schulze}, G.} \&
  \bibinfo{author}{{Pascual}, J.~I.}
\newblock \bibinfo{title}{{Competition of Superconducting Phenomena and Kondo
  Screening at the Nanoscale}}.
\newblock \emph{\bibinfo{journal}{Science}} \textbf{\bibinfo{volume}{332}},
  \bibinfo{pages}{940} (\bibinfo{year}{2011}).

\bibitem{Ruby2015}
\bibinfo{author}{Ruby, M.} \emph{et~al.}
\newblock \bibinfo{title}{Tunneling processes into localized subgap states in
  superconductors}.
\newblock \emph{\bibinfo{journal}{Phys. Rev. Lett.}}
  \textbf{\bibinfo{volume}{115}}, \bibinfo{pages}{087001}
  (\bibinfo{year}{2015}).

\bibitem{PilletNP2010}
\bibinfo{author}{Pillet, J.-D.} \emph{et~al.}
\newblock \bibinfo{title}{Andreev bound states in supercurrent-carrying carbon
  nanotubes revealed}.
\newblock \emph{\bibinfo{journal}{Nature Physics}}
  \textbf{\bibinfo{volume}{6}}, \bibinfo{pages}{965--969}
  (\bibinfo{year}{2010}).

\bibitem{Deacon2010a}
\bibinfo{author}{{Deacon}, R.~S.} \emph{et~al.}
\newblock \bibinfo{title}{{Tunneling Spectroscopy of Andreev Energy Levels in a
  Quantum Dot Coupled to a Superconductor}}.
\newblock \emph{\bibinfo{journal}{Phys. Rev. Lett.}}
  \textbf{\bibinfo{volume}{104}}, \bibinfo{pages}{076805}
  (\bibinfo{year}{2010}).

\bibitem{Grove-Rasmussen2009}
\bibinfo{author}{Grove-Rasmussen, K.} \emph{et~al.}
\newblock \bibinfo{title}{{Superconductivity-enhanced bias spectroscopy in
  carbon nanotube quantum dots}}.
\newblock \emph{\bibinfo{journal}{Phys. Rev. B}} \textbf{\bibinfo{volume}{79}},
  \bibinfo{pages}{134518} (\bibinfo{year}{2009}).

\bibitem{Lee2014a}
\bibinfo{author}{Lee, E. J.~H.} \emph{et~al.}
\newblock \bibinfo{title}{Spin-resolved {Andreev} levels and parity crossings
  in hybrid superconductor-semiconductor nanostructures}.
\newblock \emph{\bibinfo{journal}{Nat. Nanotechnol.}}
  \textbf{\bibinfo{volume}{9}}, \bibinfo{pages}{79--84} (\bibinfo{year}{2014}).

\bibitem{ChangPRL2013}
\bibinfo{author}{{Chang}, W.}, \bibinfo{author}{{Manucharyan}, V.~E.},
  \bibinfo{author}{{Jespersen}, T.~S.}, \bibinfo{author}{{Nyg{\aa}rd}, J.} \&
  \bibinfo{author}{{Marcus}, C.~M.}
\newblock \bibinfo{title}{{Tunneling Spectroscopy of Quasiparticle Bound States
  in a Spinful Josephson Junction}}.
\newblock \emph{\bibinfo{journal}{Phys. Rev. Lett.}}
  \textbf{\bibinfo{volume}{110}}, \bibinfo{pages}{217005}
  (\bibinfo{year}{2013}).

\bibitem{Jellinggaard2016}
\bibinfo{author}{{Jellinggaard}, A.}, \bibinfo{author}{{Grove-Rasmussen}, K.},
  \bibinfo{author}{{Madsen}, M.~H.} \& \bibinfo{author}{{Nyg{\aa}rd}, J.}
\newblock \bibinfo{title}{{Tuning Yu-Shiba-Rusinov states in a quantum dot}}.
\newblock \emph{\bibinfo{journal}{Phys. Rev. B}} \textbf{\bibinfo{volume}{94}},
  \bibinfo{pages}{064520} (\bibinfo{year}{2016}).

\bibitem{Kezilebiekearxiv2017}
\bibinfo{author}{Kezilebieke, S.}, \bibinfo{author}{Dvorak, M.},
  \bibinfo{author}{Ojanen, T.} \& \bibinfo{author}{Liljeroth, P.}
\newblock \bibinfo{title}{{Coupled Yu-Shiba-Rusinov states in molecular dimers
  on NbSe$_2$}}  (\bibinfo{year}{2017}).
\newblock \eprint{ArXiv1701.03288}.

\bibitem{Nadj-Perge2014}
\bibinfo{author}{{Nadj-Perge}, S.} \emph{et~al.}
\newblock \bibinfo{title}{{Observation of Majorana fermions in ferromagnetic
  atomic chains on a superconductor}}.
\newblock \emph{\bibinfo{journal}{Science}} \textbf{\bibinfo{volume}{346}},
  \bibinfo{pages}{602--607} (\bibinfo{year}{2014}).

\bibitem{PientkaPRB2013}
\bibinfo{author}{Pientka, F.}, \bibinfo{author}{Glazman, L.~I.} \&
  \bibinfo{author}{von Oppen, F.}
\newblock \bibinfo{title}{Topological superconducting phase in helical shiba
  chains}.
\newblock \emph{\bibinfo{journal}{Phys. Rev. B}} \textbf{\bibinfo{volume}{88}},
  \bibinfo{pages}{155420} (\bibinfo{year}{2013}).

\bibitem{Schecter2016}
\bibinfo{author}{Schecter, M.}, \bibinfo{author}{Flensberg, K.},
  \bibinfo{author}{Christensen, M.~H.}, \bibinfo{author}{Andersen, B.~M.} \&
  \bibinfo{author}{Paaske, J.}
\newblock \bibinfo{title}{{Self-organized topological superconductivity in a
  Yu-Shiba-Rusinov chain}}.
\newblock \emph{\bibinfo{journal}{Phys. Rev. B}} \textbf{\bibinfo{volume}{93}},
  \bibinfo{pages}{140503} (\bibinfo{year}{2016}).

\bibitem{FulgaNJPh2013}
\bibinfo{author}{{Fulga}, I.~C.}, \bibinfo{author}{{Haim}, A.},
  \bibinfo{author}{{Akhmerov}, A.~R.} \& \bibinfo{author}{{Oreg}, Y.}
\newblock \bibinfo{title}{{Adaptive tuning of Majorana fermions in a quantum
  dot chain}}.
\newblock \emph{\bibinfo{journal}{New Journal of Physics}}
  \textbf{\bibinfo{volume}{15}}, \bibinfo{pages}{045020}
  (\bibinfo{year}{2013}).

\bibitem{SuNatComm2017}
\bibinfo{author}{{Su}, Z.} \emph{et~al.}
\newblock \bibinfo{title}{{Andreev molecules in semiconductor nanowire double
  quantum dots}}.
\newblock \emph{\bibinfo{journal}{Nature Communications}}
  \textbf{\bibinfo{volume}{8}}, \bibinfo{pages}{585} (\bibinfo{year}{2017}).

\bibitem{ShermanNatNano2017}
\bibinfo{author}{Sherman, D.} \emph{et~al.}
\newblock \bibinfo{title}{{Normal, superconducting and topological regimes of
  hybrid double quantum dots}}.
\newblock \emph{\bibinfo{journal}{Nat Nano}} \textbf{\bibinfo{volume}{12}},
  \bibinfo{pages}{212--217} (\bibinfo{year}{2017}).

\end{thebibliography}

\end{document}